\documentclass[aoas,nameyear,seceqn,dvips]{arximspdf}
\usepackage{stfloats}
\usepackage{graphicx}
\usepackage{breakurl}
\doi{10.1214/10-AOAS328}
\volume{4}
\issue{3}
\pubyear{2010}
\firstpage{1234}
\lastpage{1255}

\makeatletter
\fnbelowfloat
\makeatother

\begin{document}
\begin{frontmatter}

\title{Small area estimation of the homeless in Los~Angeles: An
application of cost-sensitive stochastic~gradient boosting\thanksref{T1}}
\runtitle{Small area estimation of the homeless in Los Angeles}
\thankstext{T1}{Supported by NSF Grant SES-0437179,
``Ensemble Methods for Data Analysis in the Behavior, Social, and Economic
Sciences.''}

\begin{aug}
\author[A]{\fnms{Brian} \snm{Kriegler}\ead[label=e1]{bkriegler@econone.com}\corref{}}
\and
\author[B]{\fnms{Richard} \snm{Berk}\ead[label=e2]{berkr@sas.upenn.edu}}
\runauthor{B. Kriegler and R. Berk}
\affiliation{Econ One Research and University of Pennsylvania}
\address[A]{Econ One Research\\
601 W. 5th Street\\
Fifth Floor\\
Los Angeles, California 90071\\USA\\
\printead{e1}} 
\address[B]{Department of Criminology\\
School of Arts and Sciences\\
University of Pennsylvania\\
483 McNeil Building \\
3718 Locust Walk\\
Philadelphia, Pennsylvania 19104-6286\\USA\\
\printead{e2}}
\end{aug}

\received{\smonth{3} \syear{2009}}
\revised{\smonth{1} \syear{2010}}

%
\begin{abstract}
In many metropolitan areas efforts are made to count the homeless to ensure
proper provision of social services. Some areas are very large, which makes
spatial sampling a viable alternative to an enumeration of the entire terrain.
Counts are observed in sampled regions but must be imputed in
unvisited areas.
Along with the imputation process, the costs of underestimating and
overestimating
may be different. For example, if precise estimation in areas with
large homeless c
ounts is critical, then underestimation should be penalized more than
overestimation
in the loss function. We analyze data from the 2004--2005 Los Angeles
County homeless study using an augmentation of $L_1$ stochastic
gradient boosting that can weight overestimates and underestimates
asymmetrically. We discuss our choice to utilize stochastic gradient
boosting over other function estimation procedures. In-sample fitted
and out-of-sample imputed values, as well as relationships between the
response and predictors,
are analyzed for various cost functions. Practical usage and policy
implications of these results are discussed briefly.
\end{abstract}

%
\begin{keyword}
\kwd{Homeless}
\kwd{boosting}
\kwd{statistical learning}
\kwd{costs}
\kwd{imputation}
\kwd{quantile estimation}
\kwd{small area estimation}.
\end{keyword}\vspace*{-2pt}

\end{frontmatter}
%

\section{Introduction}\label{section:introduction}

Dating as far back as the 1930s, homelessness has been a visible,
public issue in the United States [\citet{rossi1}]. At least over the
past decade, the homeless problem has been underscored due to the rise
in unemployment and foreclosures. In the 2010 census, there are no
plans to perform street counts, thereby making it challenging for
stakeholders (e.g., homeless service advocates and selected government
agencies) to estimate the magnitude of the necessary social resources.
This is especially difficult in large metropolitan areas because the
homeless are often dispersed due to the changing availability of
homeless services, commercial development and the government's homeless
criminalization practices [\citet{berkbrownzhao}]. Areas needing
these services are literally ``moving targets.'' Adequate spatial
apportionment of homeless-related resources requires a great deal of
local information that is oftentimes prohibitively expensive to obtain.

In a typical census design, people are contacted through their place of
residence. With the possible exception of individuals living on private
property, the homeless will not be found using this design [\citet
{rossi1}]. An alternative approach is to locate homeless individuals in
temporary shelters or while they are receiving services (e.g., meals)
from public and private agencies. It is widely known, however, that a
large number of the homeless still will not be found this way because
many do not use these services. Therefore, it is common for enumerators
to canvas geographical areas and to count the homeless as they find
them. Some metropolitan areas are very large, making spatial sampling a
viable substitute to a full canvasing. One trades a reduction in the
burden of data collection in exchange for the need to impute homeless
counts for locales not visited by enumerators.

Estimation and imputation raise the issue of how best to represent the
cost of underestimation relative to overestimation (``cost function'').
The apportionment of homeless-related resources depends, at least in
part, on the estimated size of the local homeless population. Some
stakeholders, such as homeless service providers, are more troubled by
the prospect of numbers that are too small rather than too large. This
is especially true in areas where homeless counts are high, in which
undercounting may carry serious consequences. Other stakeholders, such
as elected city officials faced with budget constraints, may have the
opposite preference. In general, one needs the flexibility to penalize
overestimation and underestimation distinctly.

The homeless problem is especially serious in Los Angeles, which has a
large homeless population and consists of specific areas with very
densely populated homeless encampments [\citet{berkkrieglerylvisaker}]. These encampments can be a nuisance to
local commerce and can compound the demand, for example, for police and
hospital services [\citet{harcourt}]. One such area is ``Skid Row''
[\citet{magnanoblasi}], located just outside downtown Los Angeles. Historically, this area has been marked by high crime rates in terms of drug
markets, robberies, vandalism and prostitution, as well as drug and
alcohol abuse [\citet{lopez}].\footnote{In 2005, the Los Angeles Police
Department tested a pilot program, called ``Safer Cities Initiative''
(SCI), which was designed to target specific geographical crime ``hot
spots'' [\citet{wilsonkelling}; \citet{bratton}]. Part of this program
entailed reducing the density of homeless encampments. A full-scale
version of SCI began in September 2006 [\citet{berkmacdonald}].}
Individuals (especially the homeless) who spend significant amounts of
their time in public areas of such locales have higher victimization
rates than those who reside outside these areas [\citet{koegel}; \citet{kushel}]. In short, the set of public and private resources
dependent on the homeless population extends beyond the services
dedicated to the homeless' physical and mental health (e.g., soup
kitchens, shelters, affordable housing, etc.).

In 2004--2005, the Los Angeles Homeless Services Authority (LAHSA)
estimated the homeless population in Los Angeles County as the
aggregate of people who were living on the streets, in shelters or who
were ``nearly homeless'' (i.e., homeless people living on private
property with the consent of its residents). At any given time,
shelters cater to just a fraction of the local homeless population;
consequently, locating and estimating the street count was a daunting
task.\footnote{Homeless people were paid \$10 per hour to help the
field researchers identify locations in which the homeless could be
found. Presumably, this helped address the problem of finding ``hidden
homeless'' [\citet{rossi1}].} It would have been prohibitively costly to
canvas the entire county, which covers over 4000~square miles,
includes 2054 census tracts, and is the most populous county in the
United States.

A stratified spatial sampling of census tracts called for two steps.
First, tracts believed to have large numbers of homeless people were
visited with probability~1. There were 244 tracts of this nature, known
as ``hot tracts.'' The second step was to visit a stratified random
sample of tracts from the population of nonhot tracts. The strata were
the county's eight Service Provision Areas (SPAs), and the number of
tracts drawn from each stratum was proportional to the number of tracts
assigned to each SPA. In all, there were 265 tracts in the stratified
random sample, leaving 1545 tracts' counts to be imputed.\footnote{This is a ``small area estimation'' analysis. \citet{rao} defines a
domain, or area, as ``small'' if ``the domain-specific sample is not
large enough to support direct estimates of adequate precision.'' In
the context, homeless counts in the 265 randomly sampled tracts were
used to impute the numbers of homeless people in unvisited tracts and
ultimately the entire county.} In that analysis, the cost function was
symmetric, and emphasis was placed on estimating the homeless
population within each SPA, for various aggregations (e.g., cities),
and for the entire county [\citet{berkkrieglerylvisaker}]. Almost
certainly, symmetric costs are insufficiently responsive to the policy
needs of local stakeholders because both actual and imputed counts can
vary dramatically.

In this paper we re-analyze the Los Angeles data of 1810 nonhot tracts
using stochastic gradient boosting [\citet{friedmanboost2}] subject to
an asymmetrically weighted absolute loss function. We focus on
evaluating the relationship between homeless counts and covariates in
visited tracts and imputing the counts in unvisited tracts. By boosting
a cost-sensitive loss function, we are able to respond to the cost
functions of various stakeholders and focus on a particular region of
the conditional response. Depending on which cost function is applied,
widely varying fitted and imputed values can follow. We also explore
how different regions of the conditional response are related to the
predictors. We show that it can be practical and instructive to employ
asymmetric costs when using boosting for function estimation and imputation.

The remainder of this paper consists of five sections plus an \hyperref[append]{Appendix}.
Section~\ref{section:data} includes a description of the Los Angeles
County homeless and census data. In Section~\ref
{section:function_estimation} we provide an overview of stochastic
gradient boosting and a literature review on cost-sensitive estimation
procedures. Our analysis of the homeless data, which includes
comparisons between fitted and observed counts, imputed counts, and
model diagnostics, is in Section~\ref{section:analysis}. Section~\ref
{section:discussion} includes a discussion on how our proposed
methodology and analysis can have a profound effect on policy-making
decisions. In Sections~\ref{section:analysis} and~\ref
{section:discussion} we stress the results based on models that place
heavier penalties on underestimating, as this represents what
stakeholders would likely employ to ensure proper allocation of
homeless-related services. We conclude the paper in Section~\ref
{section:conclusion}, in which we mention some aspects of
cost-sensitive statistical learning to be explored. In the \hyperref[append]{Appendix} we derive the functional forms for the deviance,
initial value, gradient and terminal node estimates when employing
boosting subject to asymmetrically weighted absolute loss.

\section{Data description}\label{section:data}
In the 2004--2005 Los Angeles homeless study, \citet
{berkkrieglerylvisaker} considered the use of dozens of predictors in
the estimation process.\footnote{In that study, fitted and imputed
counts were obtained using random forests [\citet{breimanrf}].} The 10
predictors in Table~\ref{tab:table_variables} were relatively important
to fitting the conditional distribution of street counts, capturing
information about each tract's geographical location, land usage,
socioeconomic information and ethnic demographic data. With the
exception of median household income and planar coordinates, all other
covariates are presented in terms of percentages. While street counts
were obtained only in sampled tracts, predictor values were available
for all of the county's tracts.

\begin{table}[b]
\caption{Names and descriptions of variables in Los Angeles County
homeless data set}
\label{tab:table_variables}
\begin{tabular*}{\textwidth}{@{\extracolsep{\fill}}ll@{}}
\hline
 & \textbf{Description} \\
\hline
Response name&\\
\quad \textit{StTotal} & Homeless street count \\
[3pt]
Predictor name &  \\
\quad \textit{Commercial} & \% of land used for commercial purposes\\
\quad \textit{Industrial} & \% of land used for industrial purposes\\
\quad \textit{MedianHouseholdIncome} & Median household income\\
\quad \textit{PctMinority} & \% of population that is non-Caucasian\\
\quad \textit{PctOwnerOcc} & \% of owner-occupied housing units\\
\quad \textit{PctVacant} & \% of unoccupied housing units\\
\quad\textit{Residential} & \% of land used for residential purposes\\
 \quad \textit{VacantLand} & \% of land that is vacant\\
\quad \textit{XCoord} & Planar longitude\\
\quad \textit{YCoord} & Planar latitude\\
\hline
\end{tabular*}
\end{table}

Looking ahead to Section~\ref{section:analysis}, none of our models are
intended to necessarily suggest causal relationships. We utilized
predictor information described in Table~\ref{tab:table_variables}
primarily to estimate the conditional distribution between \textit
{StTotal} and each covariate and to construct sensible fitted and
imputed street counts. Whether the predictors are causally related to
homeless counts is at best a secondary concern.

The distribution of \textit{StTotal} is highly unbalanced. 75 percent of the
observed counts are less than 28 people, and 22 of the 265 tracts have
at least 50 homeless, of which 11 have over 100 homeless (Min $=$ 0, Q1
$=$
4, Median $=$ 12, \mbox{Mean $=$ 21.6}, Q3 $=$ 27, Max $=$ 282). To ensure adequate
local resources, stakeholders such as police departments and homeless
shelter advocates may place heavy emphasis on accurately estimating the
counts in areas that have large homeless populations (e.g., over 100
people). If so, one is willing to trade overall accuracy for a better
fit in the right tail of the street count distribution, and
underestimates are more costly than overestimates. For policy purposes,
resources may still be adequate in an area with a predicted count of 30
people when in fact the count is 50. However, if the prediction is 30
and the actual count is 150, there may well be a severe shortage of
local resources.

\section{Estimating the conditional distribution}\label
{section:function_estimation}
Let $Y$ be a set of real response values, $X$ be a vector of one or
more real predictor variables ($1,\ldots,P$), and $f(x_i)$ be a fitting
function for observation $i$ ($i = 1,\ldots, N$). We seek to minimize
some loss function, $\Psi$, to fit the conditional response
distribution, $G(Y | X = x)$:
%
\begin{equation}\label{eq3.1}
G(Y | X = x) = \arg\min_{f} E\{\Psi(Y, f(x))\}.
\end{equation}
We could minimize the $L_1$ loss so that the estimate is
%
\begin{equation}\label{eq:absoluteloss_eq0}
G_{L_{1}}(Y | X=x) = \arg\min_{f} E\{|Y - f(x)|\},
\end{equation}
in which overestimating and underestimating
the response are weighted symmetrically, and $\hat{f}$ is the median of
$Y$. But if underestimating and overestimating are not equally costly,
then the loss criteria needs to be \textit{asymmetric}. Let $L_1(\alpha
)$ be the absolute loss function that weights underestimates by $\alpha
$ and overestimates by $1-\alpha$, where $0 \leq\alpha\leq1$. Then
$G_{L_1(\alpha)}(Y |X=x)$ is defined as
%
\begin{eqnarray}\label{eq3.3}
&&G_{L_{1}(\alpha)}(Y | X=x)\nonumber\\
&&\qquad  = \arg\min_{f} E \bigl\{ \alpha|Y - f(x)|
\cdot I\bigl(Y > f(x)\bigr)\\
&&\hspace*{77pt}{}   +
  (1-\alpha)|Y-f(x)| \cdot I\bigl(Y \leq f(x)\bigr)\bigr\},\nonumber
\end{eqnarray}
where $I(Y > f(x))$ and $I(Y \leq f(x))$ are mutually exclusive
indicator variables. For each $i=1,\ldots,N$, if $y_i$ is
underestimated, then the former equals 1 and the latter equals 0.
Conversely, if $y_i$ is estimated perfectly or is overestimated, then
these binary values are reversed. Note that $G_{L_1(\alpha)}$ reduces
to $G_{L_1}$ when $\alpha=0.5$.

In general, $\hat{f}(x)$ from equation~(\ref{eq3.3}) is the quantile of $Y$, which
exhibits a straightforward translation between the cost function (or
``cost ratio'') and descriptions of the response distribution. For
example, a 3 to 1 cost ratio implies that underestimating is three
times as costly as overestimating, the ratio of underestimates to
overestimates will be 3 to 1, and $\hat{f}$ is the $3/(3+1) \times
100=75$th percentile of~$Y$. If instead the cost ratio is
less than 1 to 1, then $\hat{f}$ is less than the median of~$Y$.
Henceforth, we refer to $\alpha/ (1-\alpha)$ as the cost ratio.

\subsection{Stochastic gradient boosting: An overview}\label
{section:boosting_background}
Stochastic gradient boosting
[\citet{friedmanboost2}] is a recursive, nonparametric procedure that
has become one of the most popular machine learning algorithms among
statisticians. It exhibits extraordinary fitting flexibility, as it can
handle any differentiable and minimizable loss function. It can handle
and produce highly complex functional forms, and there is growing
evidence that it outperforms competing procedures (e.g., bagging [\citet
{breimanbagging}], splines, CART [\citet{breimancart}] and parametric
regression) in terms of prediction error [\citet{friedmanboost1}; \citet{buhlmannyu};
\citet{madigan}], provided that one utilizes reasonable tuning
parameters.\footnote{This is especially true when the number of
predictors is large [\citet{buhlmannyu}].} Shortly after \citet
{friedmanboost1} introduced gradient boosting, \citet{friedmanboost2}
augmented the algorithm by taking a random sample of observations at
each iteration, thereby creating the \textit{stochastic} gradient
boosting machine. This additional feature to the algorithm resulted in
marked reduction in bias and variance. Given stochastic gradient
boosting's success at estimating the center of $Y|X$, one may deduce
that it also performs well at estimating other regions of the
conditional response distribution.

The stochastic gradient boosting algorithm in its most general form is
provided below\footnote{Our augmentation of stochastic gradient boosting and data analysis
were conducted using \texttt{gbm} in \texttt{R} [\citet{ridgeway2}].
We found four boosting libraries in \texttt{R} in addition to \texttt{gbm}:
\texttt{ada} [\citet{culp}; \citet{culpb}], \mbox{\texttt{GAMBoost}} [\citet{bindera}], \texttt{gbev} [\citet{sexton}] and
\texttt{mboost} [\citet{hothornR}]. The respective maintainers of these packages are Mark
Culp, Harald Binder, Joe Sexton
 and Torsten Hothorn.} [\citet{friedmanboost2}; \citet{ridgeway2};
\citet{berkdm}]:

\begin{enumerate}
\item
Initialize $\hat{f}(x)$ to the same constant value across all observations,
$\hat{f}_0(x)=\arg\min_{\rho_0}\sum_{i=1}^{N}\Psi(y_i, \rho_0)$.
\item
For $t$ in $1,\ldots,T$, do the following:
\begin{enumerate}[(c)]
\item[(a)] For $i = 1,\ldots,N$, compute the negative gradient as the
working response:
\begin{eqnarray*}
z_{ti} &=&-\biggl[\frac{\partial\Psi(y_i, f_{t-1}(x_i))}{\partial
f_{t-1}(x_i)}\biggr]_{f_{t-1}(x_i)=\hat{f}_{t-1}(x_i)}.
\end{eqnarray*}
\item[(b)]
Take a simple random sample without replacement of size $N^\prime$ from
the data set with $N$ observations.
\item[(c)]
Fit a regression tree with $K_t$ terminal nodes, $g_{t}(x) = E(z_{t}|x)$
using the randomly selected observations.
\item[(d)]
Compute the optimal terminal node estimates,
$\rho_{1_t},\ldots,\rho_{K_t}$, as
\begin{eqnarray*}
\rho_{k_t} = \arg\min_{\rho_{k_t}} \sum_{x_i \in S_{k_t}} {\Psi\bigl(y_i,
\hat{f}_{t-1}(x_i) + \rho_{k_t}\bigr)},
\end{eqnarray*}
where $S_{k_t}$ is the set of $x$-values that defines terminal node
$k$ at iteration $t$.
\item[(e)]
Again using the sampled data, update $\hat{f}_{t}(x)$ as
\[
\hat{f}_{t}(x_i) \leftarrow\hat{f}_{t-1}(x_i) +
\lambda\rho_{k_t(x_i)},
\]
where $\lambda$ is the ``learning rate.''
\end{enumerate}
\end{enumerate}

In the \hyperref[append]{Appendix} we build on equation~(\ref{eq3.3}) to derive
the deviance subject to~$L_1(\alpha)$. Subsequently, we identify the
functional form of the initial value, gradient and terminal node
estimates from steps~1, 2a and 2d of the stochastic gradient boosting algorithm.

\subsection{Literature review}\label{section:litreview}
To our knowledge, the inclusion of asymmetric costs to boosting
algorithms has applied solely to classification problems. \citet{fan}
introduce an algorithm called AdaCost, a more flexible version of
\mbox{AdaBoost} [\citet{freundschapire}].\footnote{In a follow-up study of
AdaCost and other cost-sensitive variations of AdaBoost, \citet{ting}
shows that AdaCost stumbles in certain situations, and that this could
be due to the algorithm's weighting structure.} \citet{mease} propose
a boosting algorithm called JOUS-Boost, (\textbf{J}ittering and \textbf
{O}ver/\textbf{U}nder-\textbf{S}ampling). By adding small amounts of
noise to the data and weighting the probability of selection according
to each class, one can obtain different misclassification rates than if
using no jittering or unweighted sampling according to classes. \citet{berkkrieglerbaek} incorporate costs into a classification framework
using stochastic gradient boosting by specifying a threshold between 0
and 1; observations with predicted probabilities below or above the
threshold are assigned values of 0 or~1, respectively. The threshold
was established so that the ratio of misclassification errors (false
negatives to false positives) approximated the cost ratio.

In a regression context, we found three methods capable of handling
asymmetric error costs, each building on quantile estimation. If the
functional form is specifiable {a priori}, one can employ
parametric quantile regression [\citet{koenker}]. However, if the
functional form is not known, it is important and helpful to exploit
statistical learning. Then, one could apply nonparametric quantile
regression [\citet{le}]. Yet there is evidence that ensemble
procedures, such as gradient boosting, typically yield superior
bias-variance tradeoffs in comparison [\citet{buhlmannhothorn}]. \citet
{Meinshausen} introduced quantile regression forests, an augmentation
of random forests [\citet{breimanrf}]. The drawback to this method is
that the fitted and imputed values are calculated \textit{after} all of
the trees are grown using random forests. Consequently, the conditional
response function does not adapt to the cost ratio. It follows that
there are no new partial dependence plots and predictor importance
measurements (not even when employing~$L_1$, since the usual random
forests algorithm estimates the conditional mean).

Just as with parametric quantile regression, estimates based on
$L_1(\alpha)$ stochastic gradient boosting do not necessarily increase
monotonically with respect to $\alpha$.\footnote{Incidentally, quantile
regression forests does not share this feature because the quantile
estimation is performed on the distribution of each observation's
fitted values across regression trees.} Each cost function yields a
different model and fitted values that minimize the $L_1(\alpha)$ loss.
Therefore, a fitted (or imputed) count may be 30 when the cost ratio is
5 to 1 and 20 when the cost ratio is 10 to 1. With $L_1(\alpha)$
stochastic gradient boosting, our experience---both in this case
study and with other data sets---is that (i)~all (or nearly all)
fitted and imputed values tend to increase with respect to $\alpha$,
and (ii)~when decreases do occur, they tend to be small in magnitude.
We found that the use of larger terminal node sizes can reduce this
occurrence; however, for reasons we explain in Section~\ref
{section:analysis}, we purposely grew trees that potentially had small
terminal node sizes. Ultimately, we were not concerned with this ``side
effect'' because its occurrence was rare and inconsequential, and our
analysis extended beyond simply calculating fitted and imputed values.

In summary, we employed $L_1(\alpha)$ stochastic gradient boosting for
three main reasons. First, the functional form can be arrived at
inductively. Second, we have the prospect of a good bias-variance
tradeoff. Third, we can apply unequal error costs at each step of the
function estimation process so that \textit{all} of the output is
properly cost-sensitive. We found $L_1(\alpha)$ stochastic gradient
boosting to provide a formidable set of features for this case study,
though it should not be seen as a universal preference for
cost-sensitive stochastic gradient boosting in different settings.

\section{Analysis}\label{section:analysis}
Based on our discussions with key stakeholders, including people from
LAHSA and government representatives, underestimation is typically seen
to be more problematic than overestimation. The prospect of having too
few shelter beds, for instance, is more troubling than if a few beds
are open. With this in mind, our analysis emphasizes results in which
$\alpha\geq0.5$. Output based on cost functions that penalize
overestimation more heavily are also reported, primarily to demonstrate
that they are employable if one desires.

All boosting models were built using the following tuning parameters:
10 splits per tree subject to at least 5 observations per terminal node
$k_t$, a learning rate of \mbox{$\lambda=0.001$}, and a maximum of $T=6000$
trees. For stochastic gradient boosting models, we applied these same
tuning parameters along with a random sample of $N^\prime=133$
observations (i.e., a sampling fraction of 50 percent of $N=265$,
rounded to the nearest whole number). A sensible number of iterations
was determined using 10-fold cross-validation, and we found no problems
in converging on a reasonable number of trees to grow in any of our
cost-sensitive models.\footnote{For example, in the stochastic models
when the cost ratio $\alpha/(1-\alpha) \in$ \{1 to 10, 1 to 1, 10 to 1\}, the respective ``best'' numbers of iterations were 436, 1843 and
1340. Small deviations from these numbers of iterations (e.g., 1400
trees subject to a 10 to 1 ratio) yielded no substantive differences in
any results. Just as one would expect when using symmetric costs, the
cross-validation error exhibited a concave-up parabolic behavior that
tended to decrease with respect to $t$, until it reached a number of
iterations corresponding to the minimum cross-validation error. Beyond
the minimum cross-validation error iterations, the models overfit the
data [\citet{zhang}]. The key here is that these iteration estimates are
well short of $T=6000$, suggesting that we have in fact identified a
sensible number of iterations.}

Using a handful of different learning rates and sampling fractions
ranging from 0.001 to 0.01 and 35 to 75 percent, respectively, we saw
inconsequential differences in terms of street counts estimates---both fitted and imputed---and conditional distribution diagnostics,
for each $\alpha$. The same held true for models subject to 1 to 10, 1
to 5, and 1 to 1 costs. By contrast, when we employed cost ratios of 5
to 1 and 10 to 1, we learned that the number of splits and the minimum
terminal node size can have a substantial impact on point estimates.
The \texttt{gbm} library uses the inverse of the empirical distribution
to estimate quantiles, so each terminal node estimate depends on just
one value. Given the unbalanced nature of \textit{StTotal}, differences
between consecutive values in the right tail within a terminal node can
be very large. If employing a 10 to 1 cost function and a terminal node
includes 25 points, then the estimate will be the third highest value.
The use of a highly skewed cost function implies a particular interest
in estimating the handful of large response values well, yet the top
two values in this terminal node of this size will not factor into the
estimation process. To ensure that large gradients were given ample
opportunities to be terminal node estimates, we permitted large trees
and small terminal node sizes. This was facilitated by tuning the
number of splits and the minimum number of observations in each
terminal node at each iteration.\footnote{By default, in \texttt{gbm}
each tree at each iteration has one split, subject to at least 10
observations in each terminal node.}

\subsection{Fitted and imputed street counts}\label{section:predictions}
\begin{figure}

\includegraphics{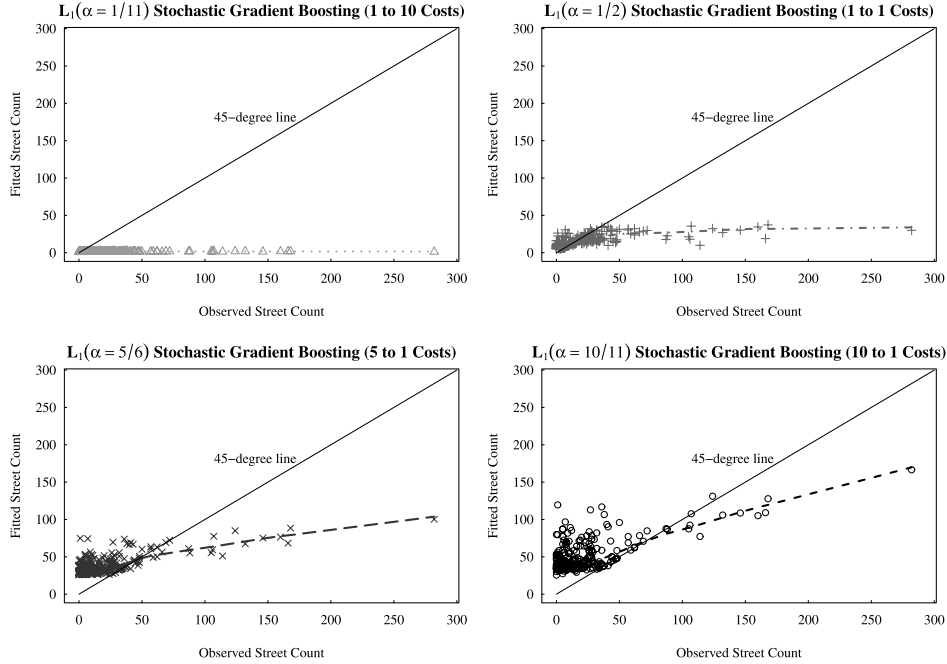}

\caption{Fitted versus observed census tract street counts
using $L_1(\alpha)$ stochastic gradient boosting.} \label{fig:obs_pred_qrb}
\end{figure}

Figure~\ref{fig:obs_pred_qrb} shows fitted versus observed street
counts for the 265 visited census tracts using stochastic gradient
boosting subject to 1 to 10, 1 to 1, 5 to 1, and 10 to 1 cost ratios
($\alpha\in\{1/11,  1/2, 5/6, 10/11 \}$, respectively). Using 1 to 1
costs ($L_1$ boosting), the magnitude of the error is less than 20
people in 232 of 265 visited census tracts. In terms of resource needs,
errors of this magnitude are likely tolerable. Conversely, among the 22
tracts with observed counts with at least 50 homeless, all of these
tracts' counts are underestimated. The maximum fitted value is
approximately 37~people, and the median error is approximately 70
people less than the true count. These large undercounts need to be
reduced substantially in order to ensure adequate local resource allocation.

Figure~\ref{fig:obs_pred_qrb} demonstrates that $L_1(\alpha)$
stochastic gradient boosting fitted values tend to increase with
respect to $\alpha$.\footnote{Of the 265 visited training data
observations, 10 observations' fitted values were lower for \mbox{$\alpha
=10/11$} than for $\alpha=5/6$. We did not consider this to be
problematic for two reasons. The largest of these differences was 4
people. Also, this was generally not a problem among tracts with very
large counts; one tract had a street count of 62, and the next highest
count was 43.} Although the overall fit worsens when the cost ratio
diverges from 1 to 1, we observe smaller errors in specific regions of
the response. Using a 10 to 1 cost ratio, just 15 out of 265 tracts are
underestimated. Among the 22 tracts with at least 50 people, the median
difference between observed and fitted counts is 1 person, and the
interquartile range is 40 people. Admittedly, most of the very large
counts are still underestimated even when using a 10 to 1 cost ratio, a
topic we will pick up again in Section~\ref
{section:discussion}.\footnote{Recognizing that it is in the nature of
all regression models to overestimate small values and underestimate
large ones, we demonstrate that the use of asymmetric costs can
alleviate the problem. As the cost ratio increases, fitted values for
tracts with large counts tend to move closer to the \mbox{45-degree} line.}

\begin{figure}

\includegraphics{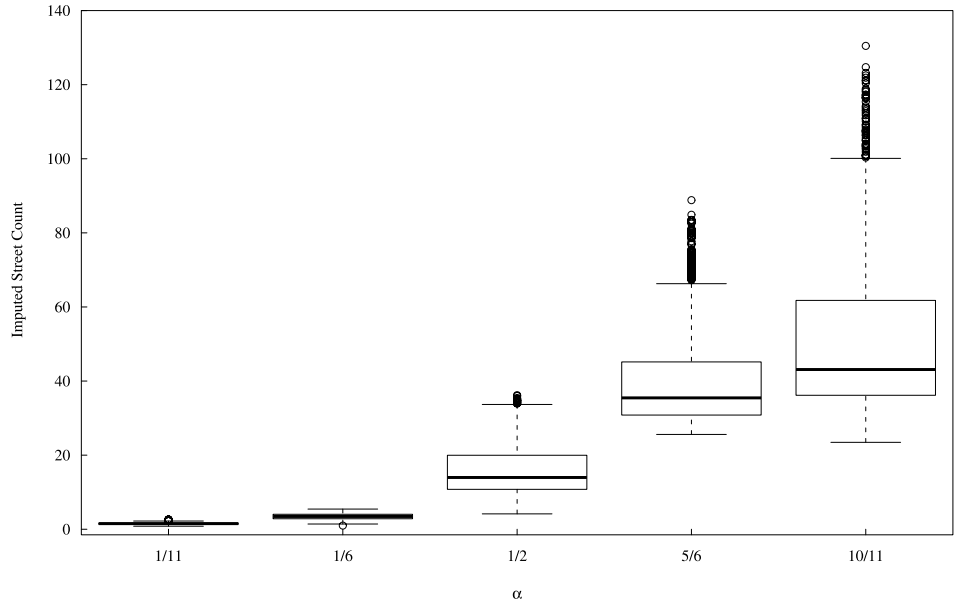}

\caption{Distribution of predicted street counts in
unvisited census tracts using $L_1(\alpha)$ stochastic gradient
boosting.} \label{fig:pred_oob_qrb}
\end{figure}

In a way, training data fitted values are irrelevant because one's
estimates of visited tracts might simply be the observed street count.
\citet{berkkrieglerylvisaker} employed this practice when they
provided estimates to LAHSA at both the tract and aggregate levels. But
provided the sampled tracts are representative of the population of all
nonhot tracts and the model does not overfit the training data, fitted
counts in Figure~\ref{fig:obs_pred_qrb} reveal how close (or far) the
unsampled tracts' imputed counts are to the true counts. Figure~\ref
{fig:pred_oob_qrb} shows the distribution of imputed counts for various
cost ratios. The distributions tend to shift upward with respect to
$\alpha$.\footnote{Of the 1545 unvisited tracts, imputed values were
higher using $\alpha=5/6$ versus $\alpha=10/11$ in 44 tracts. Over half
of these deviations were less than 2 people, and the largest deviation
was 6 people.} Using 1 to 10 and 1 to 5 costs, all tracts have imputed
counts of fewer than 5 people. Conversely, using 10 to 1 costs, we find
that 53 of 1545 tracts have imputed counts over 100 homeless people.

\begin{figure}

\includegraphics{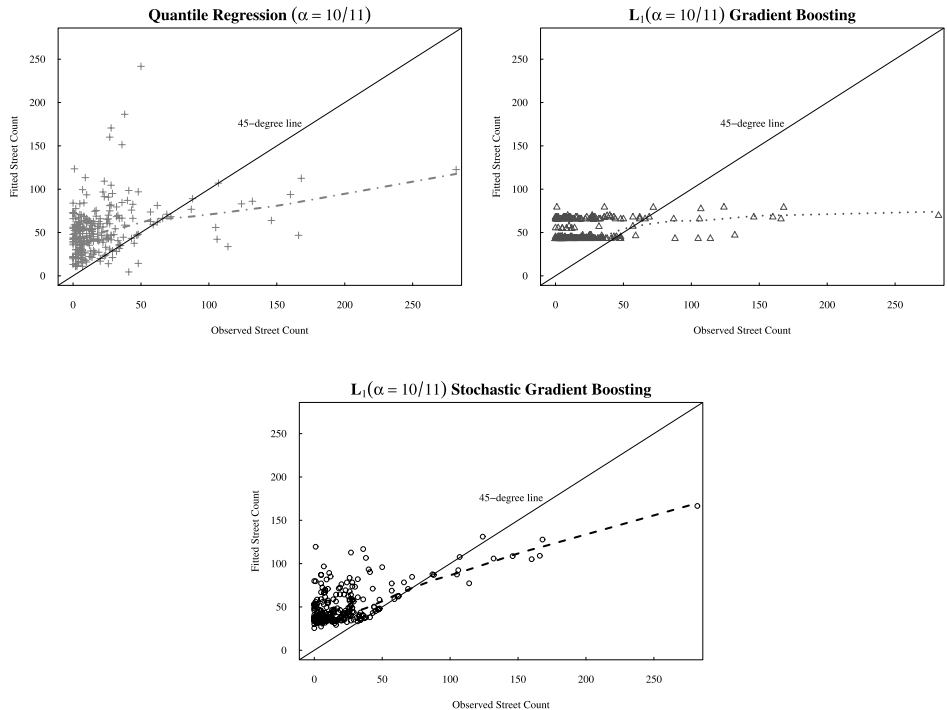}

\caption{Fitted versus observed street counts using
quantile regression, $L_1(\alpha)$ gradient boosting and $L_1(\alpha)$
stochastic gradient boosting, subject to a 10 to 1 cost ratio ($\alpha
=10/11$).} \label{fig:pred_obs_qrb_qr}
\end{figure}

Recognizing that portions of our analysis will be data set specific,
one may also be interested in how $L_1(\alpha)$ boosting performs
relative to other cost-sensitive methods. Figure~\ref
{fig:pred_obs_qrb_qr} shows fitted versus observed street counts using
stochastic and nonstochastic gradient boosting, and parametric quantile
regression, subject to a 10 to 1 cost function.\footnote{Parametric
quantile regression was performed using the \texttt{quantreg} library
in \texttt{R}, maintained by Roger \citet{koenkerr}.} All three
methods have a substantial number of overestimates, which is to be
expected given the cost ratio of choice. Among tracts with at least 50~homeless people observed, $L_1(\alpha)$ stochastic gradient boosting
performs noticeably better than the other two methods in terms of bias
and variance. Nonstochastic gradient boosting exhibits a median
deviation of 35 people underestimated and an IQR of 77 people. Quantile
regression's median deviation and IQR are 7 and 63 people, respectively.

\subsection{Conditional distribution diagnostics}\label
{section:model_diagnostics}
With 10 predictors, a highly unbalanced response distribution and
abrupt spatial variation in the data, the boosted models' conditional
distribution diagnostics are practical and necessary to understanding
relationships between the response and the predictors. Since the cost
function is built into each step of $L_1(\alpha)$ boosting, partial
plots and variable importance measures can be examined in the same
manner as when employing $L_1$ boosting. These results are especially
important if stakeholders are inclined to give causal interpretations
to the associations.

\begin{figure}

\includegraphics{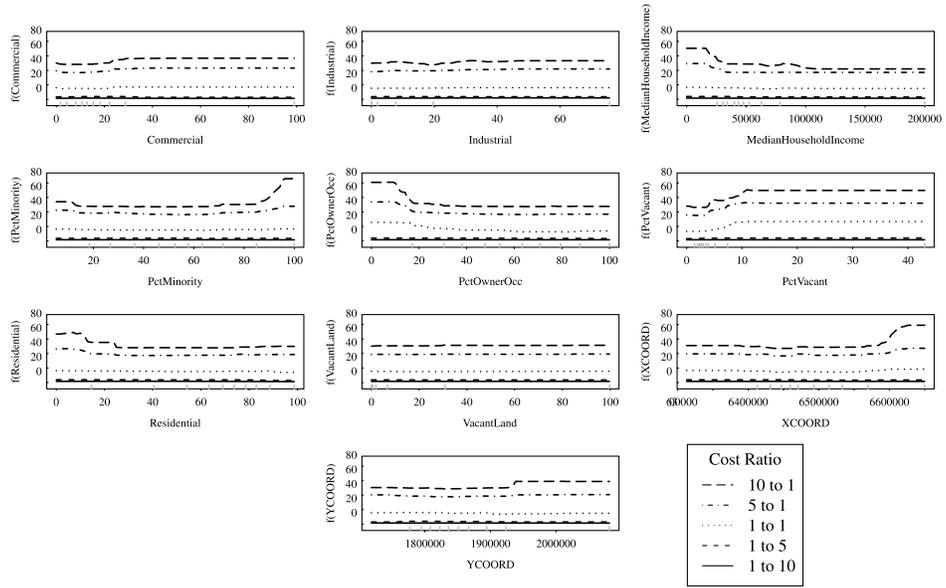}

\caption{Partial dependence plots from $L_1(\alpha)$
stochastic gradient boosting.}\label{fig:partials1}
\end{figure}

\begin{figure}

\includegraphics{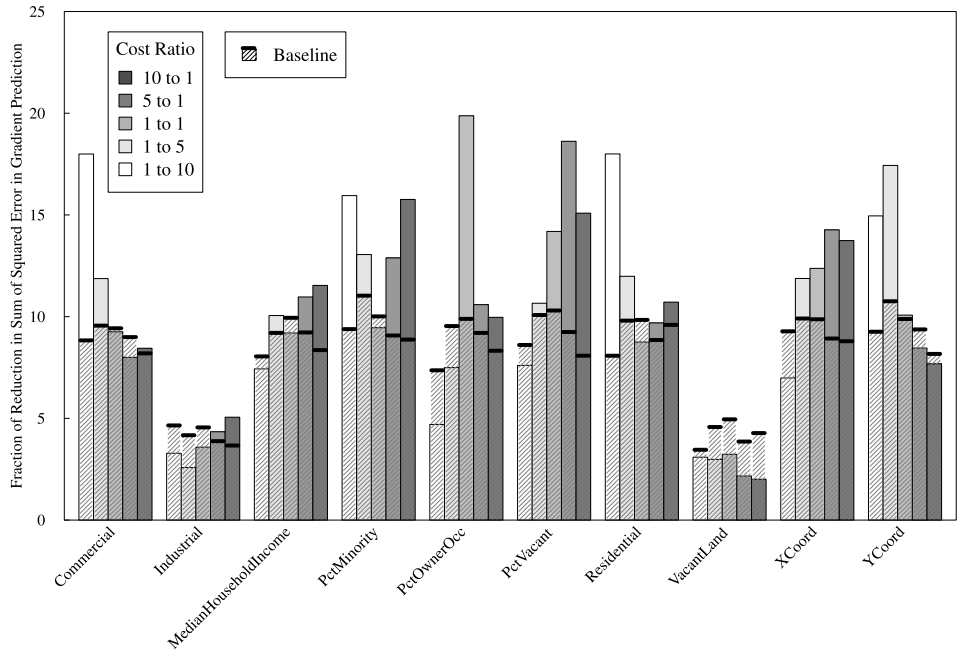}

\caption{Variable importance from $L_1(\alpha)$ stochastic
gradient boosting.} \label{fig:importance}
\end{figure}

One may assume that the partial relationships between the response and
each predictor exhibit similar directional behavior and are nothing
more than vertical shifts in the conditional response's magnitude. An
analogous argument might be made regarding variable importance: if a
predictor is important using symmetric costs, then perhaps the same is
true using asymmetric costs. If these inferences are correct,
cost-sensitive partial and predictor importance plots are less
critical. Yet Figures~\ref{fig:partials1} and~\ref{fig:importance}
demonstrate that predictors' relationships with the response are not
necessarily the same across cost ratios, underscoring the need to
examine the conditional distribution diagnostics for each cost ratio of
interest.

\subsubsection{Partial relationships}\label{section:partial}
To show partial relationships between the response and each predictor,
\citet{friedmanboost1} describes a weighted tree traversal method to
``integrate out'' all predictor variables, excluding the predictor(s)
of interest [see also \citet{ridgeway2}]. Figure~\ref{fig:partials1}
shows partial relationships between the response and each predictor for
five different cost ratios. Since each of the predictors exhibits real
values, each partial relationship is shown using a two-dimensional
smoother.\footnote{The \texttt{gbm} library estimates the partial
response at equally-spaced values (by default, 100) spanning the range
of the predictor but independent of the predictor's empirical density.
As a result, decile rugs are shown at the bottom of each plot for each
corresponding predictor to better understand the distribution of each
predictor. For example, the vacancy rate is 33 percent for one tract,
43 percent for another tract and less than 20 percent for all other
tracts. For \textit{PctVacant} greater than 20 percent, it is difficult
to determine the extent to which these partial smoothers are robust
because they are based on so few points.} For cost ratios of 1 to 10
and 1 to 5, all of the partial relationships are nearly flat, a result
consistent with the small variation in tract-level estimates reported
in Figures~\ref{fig:obs_pred_qrb} and~\ref{fig:pred_oob_qrb}. Using
symmetric $L_1$ boosting, street counts increase with respect to \textit
{PctVacant} between 0 and 10 percent, and street counts decrease with
respect to \textit{PctOwnerOcc} between 20 and 60 percent.
Pragmatically, all other partial relationships are close to null.

When underestimating \textit{StTotal} is more costly, the conditional
response can vary substantially with respect to several other
predictors in addition to the housing vacancy rates and the fraction of
owner-occupied units. For example, using a 10 to 1 cost function,
street counts are indifferent to \textit{PctMinority} until
approximately 90 percent, but increase substantially between 90 and 100
percent. Street counts decrease in a stepwise manner with respect to
\textit{MedianHouseholdIncome}; we see plateaus for incomes between \$0
and \$15,000, \$30,000 to \$75,000, and \$100,000 and above.

\subsubsection{Variable importance}\label{section:varimp}

One may be interested in identifying which predictors are ``important''
to fitting the conditional response for various cost ratios. One
measure of variable importance is the reduction in loss attributed to
each predictor. \citet{friedmanboost1} and \citet{ridgeway2} define
the ``relative influence'' as the empirical reduction in squared error
in predicting the gradient across all node splits on predictor $j$,
divided by the total reduction in error across all splits.

Even if the response and predictor $j$ are completely unrelated, it is
still possible for the predictor to be selected to split a regression
tree node. Provided there is at least one split on predictor $j$, the
empirical influence will not be zero. How then, does one know the
extent to which a predictor's influence is by chance? Along the same
lines as in random forests [\citet{breimanrf}], in which importance is
computed by shuffling each predictor in turn and comparing the change
in error, we employed the following steps to estimate each predictor's
``baseline relative influence'':

\begin{enumerate}
\item For a given predictor \textit{p}, randomly permute the values.
Keep all other predictors' values as is.
\item Construct a boosted model using the modified data in step 1 and
compute the relative influence for the shuffled predictor. Apply the
same tuning parameter settings and means for estimating a sensible
number of iterations.
\item Repeat steps 1 and 2 many times, each time computing the
relative influence of the shuffled predictor.\footnote{For $\alpha\in
\{1/11, 1/6, 1/2, 5/6, 10/11\}$, we repeated steps 1 and 2 50 times
per predictor.}
\item Compute the baseline relative influence as the average relative
influence from steps 1--3.
\item Repeat steps 1--4 for each predictor in turn.
\end{enumerate}

Figure~\ref{fig:importance} shows each predictor's empirical and
baseline relative influence values subject to five different cost
ratios. If a predictor's baseline relative influence (denoted by a
thick black line and the diagonally shaded area) is larger than its
empirical influence, this suggests that the contribution to the model
is happenstance. Just as in the partial plots, we learn that a
predictor's relative influence is not necessarily similar across cost
functions. This can be a very important practical matter insofar as
stakeholders come to accept or reject the homeless estimates depending
on whether predictors ``make sense.''

One should also be mindful of the difference between the overall
reduction in error from $t=0$---at which all estimates are equal to
the grand $\alpha$ quantile of \mbox{\textit{StTotal}}---to the ``optimal'' number
of iterations. If the total reduction in error is very small, then the
\textit{absolute} influence will be minimal. It follows that the differences
between each fitted response value and the initial constant will likely
be small as well. Under these circumstances, the relative influence
results are inconsequential. Such is the case for boosted models
subject to 1 to 10 and 1 to 5 costs. Figures~\ref{fig:obs_pred_qrb},
\ref{fig:pred_oob_qrb} and \ref{fig:partials1} suggest minimal
variation in fitted and predicted counts; substantively, the
relationships between \textit{StTotal} and each predictor are null.
Importance statistics subject to these two cost ratios are reported
primarily for demonstrative purposes.

Using symmetric costs, \textit{PctVacant} and \textit{PctOwnerOcc} are
relatively important, collectively accounting for nearly 35 percent of
the loss reduction. \textit{PctVacant} is also important when the cost
ratio is 5 to 1 or 10 to 1, along with \textit{PctMinority} and \textit
{XCoord}, and to a lesser extent \textit{MedianHouseholdIncome}. These predictors' relative influence are high compared to other
predictors' importance statistics and is well above their respective
baseline influences. Conversely, \textit{PctOwnerOcc} is much less
important when underestimation is penalized more heavily, evidenced by
its smaller relative influence and proximity to the baseline relative influence.

\section{Discussion}\label{section:discussion}
$L_1(\alpha)$ stochastic gradient boosting is a potentially useful
statistical tool for ensuring adequate allocation of services related
to the homeless. Practitioners might find it useful to build multiple
boosted models for various cost functions and examine the range of
imputed counts for a specific tract in order to make policy decisions.
Suppose a homeless service provider or local police department
considers it critical to identify tracts that have over 100 homeless
people; the former might aspire to ensure a sufficient number of beds
at the nearest shelter, and the latter may well decide to allocate
additional officers to areas with high homeless counts. Assume that a
particular tract's imputed count is 30 using 1 to 1 costs and 150 using
10 to 1 costs. Such stakeholders may insist on performing a full
enumeration in this tract because these two imputed counts have very
different resource implications. Alternatively, if the imputed counts
using these respective cost ratios are 30 and 40, a full enumeration
may not be worth the trouble because the difference is likely inconsequential.

Among the 11 tracts with over 100 homeless, stochastic gradient
boosting subject to a 10 to 1 cost ratio yields a better prediction
error than gradient boosting or parametric quantile regression. Still,
9 of the 11 tracts are underestimated, and the prediction error tends
to increase with respect to the observed count. It is reasonable to
assume that among unvisited tracts with over 100 homeless, imputed
counts will be similarly biased. In practice, one way to further reduce
this problem is by assigning larger ``population weights'' {a
priori} to training data tracts with large street counts. The
population weights increase the frequency of specific observations if
they are selected in step~2b of the algorithm described in Section~\ref
{section:boosting_background}. One assumes---and perhaps rightfully
so---that some tracts are inherently more important than others. If
larger weights are assigned to tracts with high street counts, then
fitted and imputed counts will also increase. A toy example is provided
in the~\hyperref[append]{Appendix}.

In addition to evaluating imputed counts, suppose stakeholders (e.g.,
LAHSA) want to use response-predictor relationships to determine which
unvisited tracts might require the most resources. Figure~\ref
{fig:partials1} suggests that areas with some combination of high
non-Caucasian populations, high vacancy rates, low median household
incomes and low rates of owner-occupied housing may be indicators of
high homeless populations. Based on Figure~\ref{fig:importance}, \textit
{PctVacant} and \textit{PctMinority} are especially key to identifying
areas potentially in need of services.

\section{Conclusion}\label{section:conclusion}
This case study features a number of characteristics that make the
analysis challenging. Although there are relatively few tracts with
large homeless counts, these are likely the most important tracts to
fit reasonably well---without overfitting the data---so that
unvisited tracts with potentially high counts are identified. In
addition, Los Angeles County exhibits considerable heterogeneity and
abrupt spatial changes in terms of land usage and demography. Last, the
wide range of stakeholders would likely assign various costs to
over/under-counting during the estimation and imputation processes. We
believed that a cost-sensitive ensemble statistical learning procedure
was appropriate because (i) we did not presume to understand the
underlying mechanisms of the conditional street count distribution,
(ii) we aspired to get favorable results in terms of prediction error
for specified regions of the response, and (iii) we wanted to
understand how specific regions of the conditional response were
related to the predictors. $L_1(\alpha)$ stochastic gradient boosting
allowed us to address all of these issues.

There are a handful of practical statistical issues born out of this
case study. First, one might argue that a ``cost-sensitive Poisson''
loss function is a more appropriate procedure for the homeless data
because the outcome is a count. A key issue, then, is whether $L_1$ or
$L_2$ loss is more responsive to the data imputation task at hand and
to the quality of the data. In our case, a few very large observed
counts would likely dominate the analysis under $L_2$. Whether this is
good or bad depends on the accuracy of the few very large counts and on
the policy matter of how much those large counts should be permitted to
affect the imputations. We take no strong position on either issue, but
we have concerns from past research on homeless enumerations that the
count data could contain significant error [\citet{cordray}; \citet{cowan};
\citet{rossi2}; \citet{wright}]. And, we find that boosting the $L_1(\alpha)$ loss
function incorporates cost considerations in a straightforward and
easily interpretable manner.

There is also the matter of statistical inference, a topic we glossed
over in Section~\ref{section:varimp} by estimating each predictor's
baseline relative importance. To our knowledge, statistical inference
remains a largely unsolved problem for stochastic gradient boosting and
statistical learning in general\break [\citeauthor{leeb05} (\citeyear{leeb05},
\citeyear{leeb06});
\citet{berkbrownzhao}]. We have explored the properties of a procedure that
wraps cost-sensitive boosting in bootstrap sampling cases. Although
this seems to provide some useful information on the stability of our
imputed values, we do not think it addresses the fundamental problems
identified by \citet{leeb05}.

Finally, the application of $L_1(\alpha)$ boosting brings to light the
issue of choosing the ``right'' tuning parameters, a topic explored by
\citet{mease2}. While the number of splits has been researched
extensively [e.g., \citet{schapire1};
\citet{friedmanlogistic};
\citet{buhlmannyu}; \citet{ridgeway2}], research on the impact of different terminal
node sizes is minimal thus far. Unlike estimates subject to Poisson or
Gaussian loss, which are functions of \textit{all} gradients within
each terminal node, an $L_1(\alpha)$ terminal node estimate is the
quantile of gradients residing in terminal node $k_t$. These estimates
depend on just a very local region of points and can be highly
dependent on the terminal node sizes and the way in which the quantile
is estimated [for variants of quantile estimation, see \citet
{hyndmanfan}]. The performance of $L_1(\alpha)$ stochastic gradient
boosting subject to various quantile estimation procedures remains a
topic for future research.

\begin{appendix}\label{append}
\section*{Appendix: Boosting the $L_1(\alpha)$ distribution}\label{section:derivation}

\citet{ridgeway2} specifies the boosted $L_1$ (Laplace) loss function as
%
\begin{equation}\label{eq:quantileboost_eq0}
\Psi\bigl(f_t(x_i)\dvtx  x_i \in S_{k_t}\bigr) =\biggl\{{\sum_{x_i \in S_{k_t}}
\big|{w_i}\bigl(y_i - f_t(x_i)\bigr)\big|}\biggr\} \bigg/\sum_{x_i \in
S_{k_t}}{w_i},
\end{equation}
where $w_i$ is a predetermined population weight for observation $i$
that remains constant across all iterations. Altering~(\ref
{eq:quantileboost_eq0}) to allow for unequal costs, the loss function becomes
%
\begin{eqnarray}\label{eq:quantileboost_eq1}
\Psi\bigl(f_t(x_i)\dvtx  x_i \in S_{k_t}\bigr) &=&
\Biggl\{\alpha
\mathop{\sum_{x_i \in S_{k_t}}}_{y_i > \hat{f}_t(x_i)}
       {{\big|{w_i}\bigl(y_i - \hat{f}_t(x_i)\bigr)}\big|}   \nonumber\\[-8pt]\\[-8pt]
&&\hspace*{5pt} {}+   (1 - \alpha)
\mathop{ \sum_{x_i \in S_{k_t}}}_{y_i \leq\hat{f}_t(x_i)}
     {{\big|{w_i}\bigl(y_i -
\hat{f}_t(x_i)\bigr)}\big|}\Biggr\}  \bigg /  \sum_{x_i \in S_{k_t}}{w_i},\nonumber
\end{eqnarray}
which is an asymmetrically weighted absolute loss function if $\alpha
\neq0.5$.\footnote{With this distribution, the estimate $\hat{f}$ is
in the same units as $y$; therefore, over/under-estimation are
determined by comparing the two. Estimates in some distributions, such
as Poisson, are in terms of logits and must be exponentiated to be on
the same scale as $y$.} For shorthand, denote $\Psi(f_t(x_i)\dvtx  x_i \in
S_{k_t}) = \Psi$. Then, the gradient becomes\footnote{Under the usual
$L_1$ loss function, the gradient for observation $i$ is the sign of
the difference between the observed response ($y_i$) and the predicted
value ($\hat{f}_t(x_i)$), multiplied by the population weight, $w_i$.}
%
\begin{equation}\label{eq:quantileboost_eq2}
z_{ti} = -\frac{\partial\Psi}{\partial{f_{t}(x_i)}} = \cases{
w_i \alpha\dvtx y_i > \hat{f}_{t-1}(x_i),\vspace*{2pt}\cr
-w_i (1 - \alpha) \dvtx  y_i \leq\hat{f}_{t-1}(x_i),}
\end{equation}
where the derivative is evaluated at $\hat{f}_{t-1}(x_i)$. We wish to
find the value of $\rho_{k_t}$ that minimizes $\Psi$ subject to the
loss function in~(\ref{eq:quantileboost_eq1}):
%
\begin{eqnarray}\label{eq:quantileboost_eq3}
\rho_{k_t} &=& \arg\min_{\rho_{k_t}}\Biggl\{\alpha
\mathop{\sum_{x_i \in S_{k_t}}}_{y_i > \hat{f}_{t-1}(x_i) + \rho_{k_t}}
    {{
\big|{w_i}\bigl(y_i - \bigl(\hat{f}_{t-1}(x_i) + \rho_{k_t}\bigr)\bigr)}\big|}  \nonumber\\[-8pt]\\[-8pt]
&&\hspace*{37pt}{}+  (1-\alpha)
\mathop{\sum_{x_i \in S_{k_t}}}_{ y_i \leq \hat{f}_{t-1}(x_i) + \rho_{k_t}}
     {{\big|{w_i}\bigl(y_i -
\bigl(\hat{f}_{t-1}(x_i) + \rho_{k_t}\bigr)\bigr)}\big|} \Biggr\},\nonumber
\end{eqnarray}
where $f_t(x_i)$ is the fitted value from the previous iteration, $\hat
{f}_{t-1}(x_i)$, plus the terminal node
estimate from the current iteration, $\rho_{k_t}$. Next, we
differentiate to find the value of $\rho_{k_t}$ that minimizes $\Psi$:
%
\begin{eqnarray}
\quad\qquad \frac{\partial\Psi}{\partial{\rho_{k_t}}} &=& \Biggl\{-\alpha
 \mathop{\sum_{x_i \in S_{k_t}}}_{y_i > \hat{f}_{t-1}(x_i) +\rho_{k_t}}
  {w_i} + (1 - \alpha)
\mathop{\sum_{x_i \in S_{k_t}}}_{y_i \leq \hat{f}_{t-1}(x_i) + \rho_{k_t}}
       {w_i}\Biggr\}\bigg /\sum
_{x_i \in S_{k_t}}{w_i},\\
\label{eq:quantileboost_eq4}\qquad 0 &=& -\alpha
\mathop{\sum_{x_i \in S_{k_t}}}_{y_i > \hat{f}_{t-1}(x_i) +\rho_{k_t}}
     {w_i} + (1 -
\alpha)     \mathop{\sum_{x_i \in S_{k_t}}}_{y_i \leq \hat{f}_{t-1}(x_i) + \rho_{k_t}}
     {w_i}.
\end{eqnarray}

In the right-hand side of~(\ref{eq:quantileboost_eq4}), each summation
reduces to the number of observations that are underestimated or
overestimated, respectively. Let $N_{k_t}$ denote the number of
observations in terminal node $k_t$, and let $n_{k_t}$ and $N_{k_t} -
n_{k_t}$ be the number of underestimates and overestimates in the
terminal node, respectively. For simplicity, assume that $w_i = 1$ for
all $i$. Solving for $n_{k_t}$, the location parameter is
%
\begin{equation}\label{eq:quantileboost_eq6}
n_{k_t} = {\alpha}N_{k_t}.
\end{equation}

The way in which unequal population weights affect the terminal node
estimate is worthy of a toy example. Consider terminal node $k_t$ with
5 equally-weighted observations with fitted gradients---the ``working
responses''---at $t-1$ of 0, 3, 5, 6 and 15. If we are estimating the
median, then the terminal node estimate is 5. Now suppose that prior to
constructing the boosted model, the observation with the fitted
gradient of 15 at $t-1$ was instead assigned a population weight of 3.
Then this observation's fitted gradient from $t-1$ will appear in node
$k_t$ three times, and the population-weighted median is 6.\footnote{At
present, \texttt{gbm} does not allow for unequal population weights
when employing the quantile distribution.}

By weighting the loss function according to overestimates and
underestimates, the fitted value of terminal node $k_t$ is the $\alpha$
quantile of the $N_{k_t}$ gradients. In each terminal node, there are
approximately $\alpha N_{k_t}$ and $(1-\alpha)N_{k_t}$ gradients above
and below $\rho_{k_t}$, respectively. For all $i=1,\ldots,N$,
$f_0(x_i)$ equals 0, and $\rho_0$ equals the $\alpha$ quantile of the
response variable, $y$. Therefore, the fitted value for observation $i$
after $T$ iterations, $\hat{f}_T(x_i)$, equals\footnote{Note that
$z_{ti}=0$ if observation $i$ is not randomly selected as one of the
$N^{\prime}$ observations in step~2b of the stochastic gradient
boosting algorithm described in Section~\ref{section:boosting_background}.}
%
\begin{equation}
\hat{f}_{T}(x_i) = \mathit{quantile}_{\alpha}(y) + {\lambda}\sum
_{t=1}^{T}{\mathit{quantile}_{\alpha}(z_{ti})}.
\end{equation}

Because $L_1(\alpha)$ is differentiable and there exists a solution
that minimizes this loss [\citet{hastie}], we are able to incorporate
costs into stochastic gradient boosting where the response is
quantitative, and in some sense add a distribution to those provided in
\citet{friedmanboost1}.
\end{appendix}

\section*{Acknowledgments}
We gratefully acknowledge numerous helpful discussions with Greg
Ridgeway. We would also like to thank the  Editor, anonymous
associate editor and anonymous reviewers for their constructive and
thoughtful comments on this manuscript. 

%

\printaddresses

\end{document}